\documentclass[prd,aps,twocolumn,tighten]{revtex4}
\usepackage{latexsym}
\begin{document}
\newcommand{\beq}{\begin{equation}}
\newcommand{\eeq}{\end{equation}}
\newcommand{\Prd}{Phys. Rev D}
\newcommand{\Prl}{Phys. Rev. Lett.}
\newcommand{\Plb}{Phys. Lett. B}
\newcommand{\Cqg}{Class. Quantum Grav.}
\newcommand{\Np}{Nuc. Phys.}

\title{The stability of a bouncing universe.}
\author{M.Novello and J.M.Salim}
\address{\mbox{}\\
Centro Brasileiro de Pesquisas F\'{\i}sicas,\\
Rua Dr.\ Xavier Sigaud 150, Urca 22290-180, Rio de Janeiro, RJ -- Brazil\\
E-mail: novello@cbpf.br jsalim@cbpf.br}
\date{\today}

\begin{abstract}
We investigate the stability of a spatially homogeneous and
isotropic non-singular cosmological model. We show that the
complete set of independent perturbations (the electric part of
the perturbed Weyl tensor and the perturbed shear) are regular and
well behaved functions which have no divergences, contrary to
previous claims in the literature.
\end{abstract}

\maketitle
\newpage
\section{INTRODUCTION}
\protect\label{introducao}

 The existence of singularities
appears to be a property inherent to most of the physically
relevant solutions of Einstein equations,  in particular to all
known up-to-date black hole and conventional cosmological
solutions (\cite{hawking}). In the case of black holes, to avoid
the singularity some models have been
proposed(\cite{Bardeen,Barrabes,Cabo,Mars}). These  models
nonetheless  are not exact solutions of Einstein equations since
there are no physical sources associated to them.  Many attempts
try to solve this problem by modifying general relativity
(\cite{Tseytlin,Cvetic,Horne}). More recently it has been shown
that in the framework of standard general relativity it is
possible to find spherically symmetric singularity-free solutions
of the Einstein field equations that describe a regular black
hole. The source of these solutions are generated by suitable
nonlinear vector field Lagrangians,  which  in the weak field
approximation become the usual linear Maxwell theory
(\cite{Ayon2,Ayon3,Ayon4}). Similarly in Cosmology  many
non-singular cosmological models with bounce were constructed
where the energy conditions or the validity of Einstein gravity
were violated. Such models are based on a variety of distinct
mechanisms, such as a cosmological constant (\cite{deSitter}),
non-minimal  couplings (\cite{NovelloS}), nonlinear Lagrangians
involving quadratics terms in the curvature (\cite{Mukhanov}),
modifications of the geometric structure of space-time
(\cite{NovelloS2}), quantum gravity(\cite{Nelson}), and
nonequilibrium thermodynamics (\cite{Murphy}), among others to
restrict ourselves to homogeneous and isotropic solutions. Further
investigations on regular cosmological solutions can be found in
(\cite{Veneziano}).

In a previous paper(\cite{NovelloM}) we have investigated a
cosmological model with a source produced by a nonlinear
generalization of electrodynamics and succeeded  to obtain a
regular cosmological model. The Lagrangian of our model is a
function of the field invariants up to second order. This
modification is expected to be relevant when the fields reach
large values, as is the case in the primeval era of our universe.
The model is in the framework of the Einstein field equations and
the bounce is possible because the singularity theorems
(\cite{Hawking}) are circumvented  by the appearance of a negative
pressure (although the energy density is positive definite).
Recently some papers started a detailed investigation of the
transition from contraction  to expansion in the bounce of several
models (\cite{Cartier}).  In particular, in Einstein general
relativity, models with stress-energy sources constituted by a
collection of perfect fluids and Friedmann-Robertson-Walker like
geometry were examined(\cite{nelson}). The claim in that paper is
that a generic result about the behavior  of scalar adiabatic
perturbations was obtained. The result is the following:  scalar
adiabatic perturbations can grow without limit in two situations
represented by the points where the scale factor attains its
minimum value and where $\rho+p=0$.
  The first point corresponds to the moment in which the Universe passes through the bounce;
 the second  corresponds to the transition from the region where the Null Energy
 Condition (NEC) is violated to the region where it is not.
 We will show that these instabilities are not an intrinsic property
 of a model with bounce as claimed in reference\cite{nelson}
 but a consequence of the existence of a divergence already appearing
 in the background solution if the description of the source is
 made in terms of a perfect fluid. We will present a specific example of a model with bounce,
 generated by a source representing two
 non-interacting perfect fluids, that has regular perturbations in
 the  situations  described on reference \cite{nelson}.

\section{The Model} \protect\label{The
Model}


We limit our analysis to a model \cite{NovelloM} in which the
singularity of the Friedmann-Robertson-Walker (FRW) geometry is
avoided by the introduction of nonlinear corrections to Maxwell
electrodynamics. We will consider  modifications up to second
order terms in the field invariants
\begin{equation}
L = -\frac{1}{4}\,F + \alpha\,F^2 + \beta\,G^2, \label{Order2}
\end{equation}
where $ F=F_{\mu\nu}F^{\mu\nu}, G \stackrel{.}{=} 
\frac{1}{2}\eta_{\alpha\beta\mu\nu}F^{\alpha\beta}F^{\mu\nu}$,
$\alpha$ and  $\beta$ are arbitrary constants\cite{footnote2}. The
term $FG$ will not be included in order to preserve parity. The
energy-momentum tensor for nonlinear electrodynamic theories reads
\begin{equation}
\protect\label{Tmunu} T_{\mu\nu}=-4\,L_F\,F_\mu\mbox{}^\alpha
F_{\alpha\nu} + (GL_G-L)\,g_{\mu\nu},
\end{equation}
 where $L_{F}$ represents the partial derivative of the Lagrangian
with respect to the invariant $F$ and similarly for the invariant
$G$. In the early universe, matter should be identified
with a primordial hot plasma  \cite{Tajima,
Campos} and as a consequence we are led to the case in which only
the average of the squared magnetic field survives (see
\cite{NovelloM} and references cited there for details). Since the
average procedure is independent of the equations of the
electromagnetic field we can use it in the generic expression of
the energy-momentum tensor to obtain

\begin{equation}
\protect\label{Tmunubis} T_{\mu\nu} = (\rho + p)\, v_{\mu}v_{\nu}
- p\,g_{\mu\nu},
\end{equation}
where
\begin{eqnarray}
\rho_\gamma &=& \frac{1}{2} \, H^2 \,(1 - 8\,\alpha\,H^2),
\label{rho}\\[1ex]
\protect p_\gamma &=& \frac{1}{6} \,H^2 \,(1 - 40\,\alpha\,H^2).
\label{P1}
\end{eqnarray}
 The standard result of the linear Maxwell theory can be recovered by
 setting $\alpha = \beta = 0.$

 We set
for the fundamental line element
\begin{equation}
ds^2 = dt^2 - \frac{a^2(t)}{1+\epsilon r^2/4}\,
\left[dr^2+r^2\,(d\theta^2+\sin^2\theta\,d\varphi^2)\right],
\end{equation}
where $\epsilon=-1,\,0,\,+1$ hold for the open, flat (or
Euclidean) and closed cases, respectively. The Einstein's
equations and the equation of energy conservation written for this
metric become:

\begin{equation}
(\frac{\dot{a}}{a})^2 + \frac{\epsilon}{a^2}
-\frac{1}{3}\rho_{\gamma} =0, \label{dota}
\end{equation}
\begin{equation}
-2\frac{\ddot{a}}{a}-(\frac{\dot{a}}{a})^2-\frac{\epsilon}{a^2}-p_{\gamma}
= 0, \label{p}
\end{equation}
\begin{equation}
\dot{\rho_{\gamma}} + 3(\rho_{\gamma} + p_{\gamma})
\frac{\dot{a}}{a} = 0, \label{dotrho}
\end{equation}

Inserting (\ref{rho}) and (\ref{P1}) in (\ref{dotrho}) yields for
 the magnetic field:
\begin{equation}
H=H_{0} a^{-2}, \label{ha}
\end{equation}
where $H_{0}$ is an arbitrary constant. With this result the
equation (\ref{dota}) can be integrated. For the case $\epsilon
=0$ the  solution is:
\begin{equation}
a(t)^2=H_0\left[\frac{2}{3}(kc^2t^2+12\alpha)\right]^{1/2}.\label{at}
\end{equation}

 The interpretation of the source as a one component perfect
fluid in an adiabatic regime has some difficulties that are at the
origin of the instabilities found in (\cite{nelson}).  The sound
velocity of the fluid in this case is given by (\cite{Landau})
\begin{equation}
\left
(\frac{\partial{p_{\gamma}}}{\partial{\rho_{\gamma}}}\right)
=\frac{\dot{p_{\gamma}}}{\dot{\rho_{\gamma}}}
=-\frac{\dot{p_{\gamma}}}{\theta(\rho_{\gamma}+p_{\gamma})}
\end{equation}
This expression, involving only the background, is not defined at
the points where the energy density attains an extremum given by
$\theta=0$ and $\rho_{\gamma}+p_{\gamma}=0$. In terms of the
cosmological time they are determined by $t=0$ and $t=\pm
t_c=12\alpha/k c^2$. These points are well-behaved regular points
of the geometry indicating that the description of the source is
note appropriate.

This difficulty can be circumvented if we adopt another
description for the source of the model.  This can be achieved if
we  separate the part of the source related to Maxwell dynamics
from the additional non-linear $\alpha-$dependent term on the
Lagrangian. By doing this the source automatically splits in two
noninteracting perfect fluids:
\begin{equation}
T_{\mu\nu} = T^{1}_{\mu\nu} + T^{2}_{\mu\nu},
 \protect\label{ex1}
\end{equation}
where,
\begin{equation}
\protect\label{Tmunubis1} T^{1}_{\mu\nu} = (\rho_{1} + p_{1})\,
v_{\mu} v_{\nu} - p_{1}\,g_{\mu\nu},
\end{equation}
\begin{equation}
\protect\label{Tmunubis2} T^{2}_{\mu\nu} = (\rho_{2} + p_{2})\,
v_{\mu} v_{\nu} - p_{2}\,g_{\mu\nu},
\end{equation}
and
\begin{eqnarray}
\rho_{1} &=& \frac{1}{2} \, H^2
\label{rho1}\\[1ex]
\protect p_{1} &=& \frac{1}{6} \,H^2. \label{P21} \\[1ex]
\protect  \rho_{2} &=&- 4\alpha \, H^4
\label{rho1}\\[1ex]
\protect p_{2} &=& -\, \frac{20}{3}\, \alpha \,H^4. \label{P22}
\end{eqnarray}
Using the above decomposition it follows that each one of the two
components of the fluid satisfy independently equation
(\ref{dotrho}). This indicates that the source can be described by
two non-interacting perfect fluids with equation of states $ p_{1}
= 1/3 \,\rho_{1}$ and $ p_{2} = 5/3 \,\rho_{2}.$  The equation of
state for the second fluid should be understood only formally as a
mathematical device to allow for a fluid description.

\section{Gauge invariant treatment of perturbation}
\protect\label{gauge}

 In a series of papers \cite{Novello1,Novello2,Novello3}
we have established a complete and self-consistent theory to deal
with the problem of perturbations of the FRW cosmology. The very
well-known problem of the gauge dependence of the perturbations
was addressed and solved by the introduction of a complete set of
gauge invariants variables that represents direct observable
quantities. We present here a summary of this formalism in order
to fix the notation and to aim for the self consistence of this
paper.

The source of the background geometry  is represented by
two-fluids, each one having an independent equation of state
relating the pressure and the energy density  ($p_{i} =
\lambda_{i} \rho_{i},$ where the  $i=1,2$.  Following the standard
procedure we consider arbitrary perturbations that preserves each
equation of state. Thus the general form of the perturbed
energy-momentum tensor is written as
\begin{equation}
\delta T^{i}_{\mu\nu} = (1 + \lambda_{i}) \hspace{0.1cm}\delta
(\rho_{i} v_{\mu} v_{\nu}) - \lambda_{i} \delta (\rho_{i}
g_{\mu\nu}) . \protect\label{m1}
\end{equation}

The background geometry is conformally flat. Thus any perturbation
of the Weyl tensor is a true perturbation of the gravitational
field. It is convenient to represent the Weyl tensor
$W_{\alpha\beta\gamma\delta}$ in terms of its corresponding
electric and magnetic parts (these names come from the analogy
with the electromagnetic field):
\begin{equation}
E_{\alpha\beta} = - W_{\alpha\mu\beta\nu} v^{\mu} v^{\nu}
\protect\label{d4}
\end{equation}
\begin{equation}
H_{\alpha\beta} = - W^{\ast}_{\alpha\mu\beta\nu} v^{\mu} v^{\nu}.
\protect\label{d5}
\end{equation}
These variables have the advantage that since they are null in the
background, their perturbations are gauge invariant quantities
\cite{Hawking}. These definitions imply for the tensors
$E_{\mu\nu}$ and $H_{\mu\nu}$ the following properties:
$$E_{\mu\nu} = E_{\nu\mu}, \, E_{\mu\nu} v^{\mu} = 0, \,
E_{\mu\nu} g^{\mu\nu} = 0,$$ and for the magnetic tensor:
$$ H_{\mu\nu} = H_{\nu\mu},\, H_{\mu\nu} v^{\mu} = 0, \,
H_{\mu\nu} g^{\mu\nu} = 0.$$

\subsection{Some mathematical machinery}

 The metric $g_{\mu\nu}$ and the vector $v^{\mu}$ (tangent to a time-like
congruence of curves $\Gamma$) induce a projector tensor
$h_{\mu\nu}$ which separates any tensor in terms of quantities
perpendicular and parallel to $ v^{\mu}$. The  projector is
defined as
\begin{equation} {h_{\mu}}^{\lambda} \equiv
{\delta_{\mu}}^{\lambda} - v_{\mu} \hspace{0.1cm} v^{\lambda},
\protect\label{d8bis}
\end{equation}
and has the  property that
\begin{equation}
 h_{\mu\nu} h^{\nu\lambda} = {h_{\mu}}^{\lambda}.
  \hspace{0.1cm}
\protect\label{d8}
\end{equation}

 The equations of motion for the first order
perturbations are linear so it is useful to develop all  perturbed
quantities in the spherical harmonics basis. In this paper we will
limit our analysis to perturbations represented in the scalar base
defined by the equation:
\begin{equation}
h^{\mu\nu}\hat{\bigtriangledown}_{\mu}\hat{\bigtriangledown}_{\nu}Q_{(n)}=
 - m^2_{n} \, \frac{1}{a^2}\, Q_{(n)},
\end{equation}

where $m$ is the wave number and $\hat{\bigtriangledown}_{\mu}$ is
the covariant derivative in the hypersurface with normal $v^{\mu}$
and  metric $h_{\mu\nu}$. From now on we will suppress the index
$n$.

The scalar $Q$  allows us to define the associated vector
$\pi_{\mu}$ and traceless tensor $P_{\mu\nu}:$
\begin{equation}
\pi_{\mu}= \frac{a^2}{m^2}\hspace{0.1cm} {h_{\mu}}^{\nu}
\hat{\bigtriangledown}_{\nu}Q,
\end{equation}
\begin{equation}
P_{\mu\nu}= \hat{\bigtriangledown}_{\mu} \pi_{\nu}
-\frac{1}{3}h_{\mu\nu} Q.
\end{equation}
In the case of scalar perturbations the fundamental set of
equations, determining the dynamics of the perturbations are (see
the appendix):
\begin{eqnarray}
(\delta
E_{1}^{\mu\nu})^{\bullet}{h_{\mu}}^{\alpha}{h_{\nu}}^{\beta} &+& (
\delta E_{2}^{\mu\nu})^{\bullet}
{h_{\mu}}^{\alpha}{h_{\nu}}^{\beta}  + \Theta
(\delta E_{1}^{\alpha\beta}+\delta
E_{2}^{\alpha\beta})  \nonumber \\
& = & - \frac{1}{2}(\rho_{1} + p_{1})\, \delta
\sigma_{1}^{\alpha\beta}\nonumber \\
&-& \frac{1}{2}(\rho_{2} + p_{2}) \, \delta
\sigma_{2}^{\alpha\beta}
\protect\label{apb13}
\end{eqnarray}
\begin{eqnarray}
(\delta E^{1}_{\alpha\mu}&+&\delta E^{2}_{\alpha\mu})_{;\nu}
h^{\alpha\varepsilon} \hspace{0.1cm}h^{\mu\nu} = \frac{1}{3}
(\delta\rho_{1}+\delta\rho_{2})_{,\alpha} h^{\alpha\varepsilon}
\nonumber \\
&-& \frac{1}{3} \dot{\rho}_{1} \hspace{0.1cm}\delta
v_{1}^{\varepsilon} -\frac{1}{3} \dot{\rho}_{2}
\hspace{0.1cm}\delta v_{2}^{\varepsilon} \protect\label{apb16}
\end{eqnarray}
\begin{eqnarray}
\left(\delta \sigma^{1}_{\mu\nu}\right)^{\bullet}&+& \left(
\delta\sigma^{2}_{\mu\nu}\right)^{\bullet}
 + \frac{1}{3}h_{\mu\nu} (\delta {a_{1}^{\alpha}}+\delta
a_{2}^{\alpha})_{;\alpha} \nonumber \\
&-& \frac{1}{2}(\delta a^{1}_{\alpha;\beta}+\delta
a^{2}_{\alpha;\beta}) \hspace{0.1cm}{h_{(\mu}}^{\alpha}
\hspace{0.1cm}{h_{\nu)}}^{\beta}
\nonumber \\
& + & \frac{2}{3}\Theta
\hspace{0.1cm}(\delta\sigma^{1}_{\mu\nu}+\delta\sigma^{2}_{\mu\nu})
= - \delta E^{1}_{\mu\nu}- \delta E^{2}_{\mu\nu}
\protect\label{apb18}
\end{eqnarray}
\begin{eqnarray}
 - \lambda_{1}\delta \left(\rho_{,\beta} \hspace{0.1cm}{h^{\beta}}_{\mu}\right) + (1
 +\lambda_{1})\rho
 \hspace{0.1cm}\delta a^{1}_{\mu}=0
\protect\label{apb24}
\end{eqnarray}
\begin{eqnarray}
 - \lambda_{2}\delta \left(\rho_{,\beta} \hspace{0.1cm}{h^{\beta}}_{\mu}\right) + (1
 +\lambda_{2})\rho
 \hspace{0.1cm}\delta a^{2}_{\mu}= 0.
\protect\label{apb25}
\end{eqnarray}
The acceleration $a^{\mu}$, the expansion $\Theta$ and the shear
$\sigma_{\mu\nu}$ that appear in the above equations are parts of
the irreducible components of the covariant derivative of the
velocity field defined as:
\begin{equation}
a^{\mu}={v^{\mu}}_{;\nu} v^{\nu},
\end{equation}
\begin{equation}
\Theta={ v^{\mu}}_{;\mu},
\end{equation}
\begin{equation}
\sigma_{\alpha\beta}=\frac{1}{2}{h^{\mu}}_{(\alpha}{h_{)\beta}}^{\nu}
v_{\mu ;\nu}- \frac{1}{3} \Theta h_{\alpha\beta} .
\end{equation}
The expansion of the perturbations in terms of the spherical
harmonic basis is\cite{footnote3}
\begin{equation}
\delta\rho= N(t)Q,
\end{equation}

\begin{equation}
\delta v^{\mu}= V(t)h^{\mu\alpha}Q_{|\alpha},
\end{equation}

\begin{equation}
\delta a^{\mu}= \dot{V} \, h^{\mu\alpha}Q_{|\alpha},
\end{equation}

\begin{equation}
\delta E^{\mu\nu}=E(t)P^{\mu\nu},
\end{equation}

\begin{equation}
\delta \sigma^{\mu\nu}= \Sigma(t)P^{\mu\nu},
\end{equation}

\subsection{Perturbation of a bouncing universe}

After presenting the necessary formalism we shall to start the
analysis of the perturbations of the bouncing cosmological model
displayed in previous section. Using the above expansion into the
equations (cf. apendix) (\ref{apb13}), (\ref{apb16}),
(\ref{apb18}), (\ref{apb24}) and (\ref{apb25}) we obtain:
\begin{equation}
E_{1}+E_{2}=\frac{a^2}{6\epsilon+k^2}\left(N_{1}+
N_{2}-{\dot{\rho}}_{1}V_{1}-{\dot{\rho}}_{2}V_{2}\right),
\end{equation}

\begin{eqnarray}
\dot{E}_{1}+ \dot{E}_{2} &+&\frac{1}{3}
\Theta\left(E_{1}+E_{2}\right)\nonumber
\\
&=& -\left(\frac{1+\lambda_{1}}{2}\right)\rho_{1}\Sigma_{1}-
\left(\frac{1+\lambda_{1}}{2}\right)\rho_{2}\Sigma_{2},
\end{eqnarray}

\begin{equation}
\dot{\Sigma_{1}}+\dot{\Sigma_{2}}-\dot{V}_{1}-\dot{V}_{2}=-E_{1}-E_{2},
\end{equation}
\begin{equation}
-\lambda_{1}\left(N_{1}-\dot{\rho}_{1}V_{1}\right)+
\left(1+\lambda_{1}\right)\rho_{1}\dot{V}_{1}=0,
\end{equation}
\begin{equation}
-\lambda_{2}\left(N_{2}-\dot{\rho}_{2}V_{2}\right)+
\left(1+\lambda_{2}\right)\rho_{2}\dot{V}_{2}=0,
\end{equation}
These equations can be rewritten in a more convenient way as:
\begin{equation}
\dot{\Sigma_{1}}= -
\left(\frac{2\lambda_{1}(3\epsilon+k^2)}{a^2(1+\lambda_{1})\rho_{1}}+1\right)E_{1},
\end{equation}
\begin{equation}
\dot{\Sigma_{2}}= -
\left(\frac{2\lambda_{1}(3\epsilon+k^2)}{a^2(1+\lambda_{2})\rho_{2}}+1\right)E_{2},
\end{equation}
\begin{equation}
\dot{E_{1}}+\frac{1}{3}\Theta
E_{1}=-\frac{1}{2}\left(1+\lambda_{1}\right)\rho_{1}L\Sigma_{1},
\end{equation}
\begin{equation}
\dot{E_{2}}+\frac{1}{3}\Theta
E_{2}=-\frac{1}{2}\left(1+\lambda_{2}\right)\rho_{2}\Sigma_{2},
\end{equation}
As we have shown in (\cite{Novello1}), the whole set of scalar
perturbations can be expressed in terms of the two basic
variables:  $E_{i}$ and $\Sigma_{i}$. The corresponding equations
can be decoupled. The result in terms of variables $E_{i}$ is the
following:
\begin{eqnarray}
\ddot{E}_{i}&+&\frac{4+3\lambda_{i}}{3}\Theta \dot{E}_{i}+ \{\frac
{2+3\lambda_{i}}{9}\Theta^2\nonumber
\\
&-&(\frac{2}{3}+\lambda_{i})\rho_{i}\nonumber
\\
&-&\frac{1}{6}(1+3\lambda_{j})\rho_{j}
-\frac{3\epsilon+k^2)\lambda_{i}}{a^2} \} E_{i}=0.
\end{eqnarray}


Note that there is no sum in indices and $j\neq i$ in this
expression. In our case the values of $\lambda_{i}$ are
$\lambda_{i}=\left(\frac{1}{3}, \frac{5}{3}\right)$. In the first
case the equation for the variable $E_{1}$ became:
\begin{equation}
\ddot{E}_{1}+\frac{5}{3} \Theta \dot{E}_{1}+\left[\frac
{1}{3}\Theta^2-\rho_{1}-\rho_{2} -\frac{5 k^2}{3 a^2}\right]
E_{(1)}=0
\end{equation}

We should analyze the behavior of these perturbations in the
neighborhood of the points where the energy density attain an
extremum. This means not only the bouncing point but also the
point in which $\rho + p$ vanishes. Let us  start by the exam at
the bouncing  point $t=0$.

The expansion of the equation of $E_1$ in the neighborhood of the
bouncing, up to second order, is given by:
\begin{equation}
\ddot{E_1}+at\dot{E_1}+(b+b_1t^2)E_1=0\protect\label{equa1}
\end{equation}
The constant  $a$ and the parameter $b$ are defined as follows
\begin{equation}
a=\frac{5}{2t_c^2}
\end{equation}

\begin{equation}
b=-\frac{m^2}{\sqrt{6} H_0 t_c}
\end{equation}

\begin{equation}
b_1=-\frac{b}{2t_c^2}-\frac{3}{4t_c^4}
\end{equation}
Defining a new variable f as
\begin{equation}
f(t)=E_1(t)
exp{\left(+\frac{a}{4}-\frac{i}{2}\sqrt{b_1-\frac{a^2}{4}}\right)t^2},
\end{equation}
doing a coordinate transformation for time as indicated bellow
\begin{equation}
\xi=-i ( \sqrt{b_1-\frac{a^2}{4}}) t^2,
\end{equation}
We obtain the following confluent hypergeometric equation
(\cite{Abramowitz})
\begin{equation}
\xi \ddot{f}+(1/2-\xi)\dot{f}+ e f=0,
\end{equation}
where
\begin{equation}
e=\frac{i(b-a/2)}{4(b_1-a^2/4)^{1/2}}-\frac{1}{2}.
\end{equation}

 The
solution of this equation is given by:
\begin{equation}
f(t)= A \;  M(d,1/2,-i ( \sqrt{b_1-\frac{a^2}{4}}) t^2),
\end{equation}
where $A$ is an arbitrary constant and $M(d,1/2,\xi)$ confluent
hypergeometric function. The confluent hypergeometric function is
well behaved in this neighborhood and so also is the perturbation
$E_{1}(t)$given by:
\begin{eqnarray}
E_1(t)&=&Re [ A \; M(d,1/2,-i ( \sqrt{b_1-\frac{a^2}{4}})t^2)\nonumber \\
&*&exp{\left(-\frac{a}{4}+\frac{i}{2}\sqrt{b_1-\frac{a^2}{4}}\right)t^2}].
\end{eqnarray}

The perturbation $E_2$ at this same neighborhood, after the same
procedure we did before result in the following equation:
\begin{equation}
\ddot{E_2}+at\dot{E_2}+(b+b_1t^2)E_2=0\protect\label{equa2}
\end{equation}
This is the same equation we obtained for $E_1$, they differ only
by the values of the parameters $a, b$ and $b_1$ that in this case
are:
\begin{equation}
a=\frac{9}{2t_c^2}
\end{equation}

\begin{equation}
b= \frac{3}{2t^2_c}- 5\frac{m^2}{\sqrt{6} H_0 t_c}
\end{equation}

\begin{equation}
b_1=-\frac{5 m^2}{t_c^3 H_0\sqrt{6}}-\frac{5}{t_c^4}
\end{equation}
Then the solution in this case is:
\begin{eqnarray}
 E_2(t)&=&Re [ A\, M\left(d,1/2,-i ( \sqrt{b_1-\frac{a^2}{4}}\right) t^2)\nonumber \\
 &*& exp{-\left(\frac{a}{4}-\frac{i}{2}\sqrt{b_1-\frac{a^2}{4}}\right)t^2}],
\end{eqnarray}
Again the confluent hypergeometric function is well behaved in
this neighborhood and so also is the perturbation $E_{2}(t)$. At
the neighborhood of the point $t=t_c$ the equation for the
perturbation $E_1$ is given by
\begin{eqnarray}
\ddot{E}_1 + a\dot{E}_1 + \left(b+b_1
t\right)E_1=0
\protect\label{equa3}
\end{eqnarray}
where the parameters $a, b, b_1$ in this case are given by
\begin{equation}
a=\frac{5}{4t_c}
\end{equation}
\begin{equation}
b_=-\frac{3}{4t_c^2}-\frac{\sqrt{3} m^2}{6 H_0 t_c}
\end{equation}
\begin{equation}
b_1=\frac{\sqrt{3}}{4t_c^2}\left(
\frac{m^2}{3H_0}-\frac{3}{2t_c}\right)
\end{equation}

 We would like to remark that this equation is different from the
equations (\ref{equa1},\ref{equa2}) obtained in the neighborhood
of $t=0$. We proceed doing the following variable transformation:
\begin{equation}
E_1(t)=\exp{-\frac{a t}{2}} w(t)
\end{equation}
The differential equation for this new variable is
\begin{equation}
\ddot{w}+\left( b-(a/2)^2+b_1t\right)w=0
\end{equation}

 The solution
for this equation is
\begin{equation}
w(t)=  \left[w_0\, AiryAi\left(-\frac{b-(a/2)^2 + b_1
t}{b^{2\\/3}}\right)\right]
\end{equation}

The AiryAi are regular well behaved functions in this neighborhood
and so also the perturbations $E_1$. Finally we look for the
equation of $E_2$ at the neighborhood of $t=t_0$, it becomes
\begin{eqnarray}
\ddot{E}_2 + a\dot{E}_2 + \left(b+b_1 t\right)E_2=0
\end{eqnarray}
where the parameters $a, b, b_1$ in this case are given by
\begin{equation}
a=\frac{9}{4t_c}
\end{equation}
\begin{equation}
b=\frac{5}{t_c}\left(\frac{5}{4t_c}-\frac{\sqrt{3} m^2}{6 H_0
}\right)
\end{equation}
\begin{equation}
 b_1=\frac{5\sqrt{3}}{2t_c^2}\left(
\frac{1}{t_c}-\frac{m^2}{6H_0}\right)
\end{equation}
This equation differ from eq.(\ref{equa3}) only by the numerical
values of the parameters $a, b, b_1$ so we obtain the same regular
solution
\begin{equation}
E_2= Re \left( \exp{-\frac{at}{2}}\,w_0 \,AiryAi
[-\frac{b-(a/2)^2+ b_1 t}{b_1^{2/3}}] \right)
\end{equation}

\section{Conclusion}

Recently there has been a renewed  interest on nonsingular
cosmology. As a direct consequence of this some authors have
argued against these models based on instability reasons. In
(\cite{nelson}) it has been argued that a rather general analysis
shows that there are instabilities associated to some special
points of the geometrical configuration. They correspond to the
points of bouncing of the model and  maxima of the energy density,
where the description of the matter content in terms of a single
perfect fluid does not apply. In the present paper we have shown,
by a direct analysis of a specific nonsingular universe, that the
result claimed in the quoted paper does not apply to our model. We
took the example from a recent paper (\cite{NovelloM}) in which
the avoidance of the singularity comes from a non linear
electrodynamic theory.  We used the quasi Maxwellian equations of
motion ({\cite{Novello1}, \cite{Novello2}, \cite{Novello3}) in
order to undertake the analysis of the perturbed set of Einstein
equations of motion. We showed that in the neighborhood of the
special points in which a change of regime occurs, all independent
perturbed quantities are well behaved. Consequently the model does
not present any difficulty concerning its instability. This paves
the way to investigate models with bounce in more detail and to
consider them as good candidates to describe the evolution of the
Universe.

\section{Appendix A: comparison with others gauge-invariant
variables} \protect\label{previous}

FRW cosmology is characterized by the homogeneity of the
fundamental variables that specify its kinematics (the expansion
factor $\Theta$), its dynamics (the energy density $\rho$) and its
associated geometry (the scalar of curvature $R$). This means that
these three quantities depend only on the global time $t$,
characterized by the hypersurfaces of homogeneity. We can thus use
this fact to define in a trivial way 3-tensor associated
quantities, which vanish in this geometry, and look for its
corresponding non-identically vanishing perturbation. The simplest
way to do this is just to let $U$ be a homogeneous variable (in
the present case, it can be any one of the quantities $\rho$,
$\Theta$ or $R$), that is $U = U(t)$. Then use the 3-gradient
operator $^{(3)}\bigtriangledown_{\mu}$ defined by
\begin{equation}
^{(3)}\bigtriangledown_{\mu} \equiv {h_{\mu}}^{\lambda}
\hspace{0.1cm}\bigtriangledown_{\lambda} \protect\label{prev1}
\end{equation}
to produce the desired associated variable
\begin{equation}
U_{\mu} = {h_{\mu}}^{\lambda}
\hspace{0.1cm}\bigtriangledown_{\lambda} U. \protect\label{prev2}
\end{equation}

In \cite{Ellis} these quantities were discussed and its associated
evolution analysed. In the present section we will exhibit the
relation of these variables to our fundamental ones. We shall see
that under the conditions of our analysis\cite{footnote4} these
quantities are functionals of our basic variables ($E$ and
$\Sigma$) and of the background ones.

\subsection{The matter variable $\chi_{i}$}
\protect\label{matter variable}

It seems useful to define the fractional gradient of the energy
density $\chi_{\alpha}$ as \cite{Ellis}
\begin{equation}
\chi_{\alpha} \equiv \frac{1}{\rho} \hspace{0.1cm}
^{(3)}\bigtriangledown_{\alpha} \hspace{0.1cm}\rho.
\protect\label{1234}
\end{equation}

Such quantity $\chi_{\alpha}$ is nothing but a combination of the
acceleration and the divergence of the anisotropic stress. Indeed,
from the above equations it follows (in the frame in which there
is no heat flux)
\begin{equation}
\delta \chi_{i} = \frac{(1 + \lambda)}{\lambda} \hspace{0.1cm}
\delta a_{i} + \frac{1}{\lambda\rho} \hspace{0.1cm}
\delta{{\Pi_{i}}^{\beta}}_{;\beta} \protect\label{1235}
\end{equation}

>From what we have learned above it follows that this quantity can
be reduced to a functional of the basic quantities of
perturbation, that is $\Sigma$ and $E$, yielding
\begin{equation}
\delta \chi_{i} = -2 \hspace{0.1cm}\left(1 - \frac{3K}{m}\right)
\hspace{0.1cm} \frac{1}{\rho A^{2}} \hspace{0.1cm}\left(E -
\frac{\xi}{2} \Sigma\right) \hspace{0.1cm}Q_{i}.
\protect\label{1236}
\end{equation}

\subsection{The kinematical variable $\eta_{i}$}
\protect\label{kinematical variable}

The only non-vanishing quantity of the kinematics of the cosmic
background fluid is the (Hubble) expansion factor $\Theta$. This
allows us to define the quantity $\eta_{\alpha}$ as:
\begin{equation}
\eta_{\alpha} = {h_{\alpha}}^{\beta}
\hspace{0.1cm}\Theta_{,\beta}. \protect\label{1237}
\end{equation}
Using the constraint relation eq.(\ref{apb10}) we can relate this
quantity to the basic ones:
\begin{equation}
\delta \eta_{i} = -\frac{\Sigma}{A^{2}} \hspace{0.1cm}\left(1 -
\frac{3K}{m} \right) \hspace{0.1cm}Q_{i}. \protect\label{1238}
\end{equation}

\subsection{The geometrical variable $\tau$}
\protect\label{the geometrical variable}

We can choose the scalar of curvature $R$ which depends only on
the cosmical time $t$ like $\rho$ and $\Theta$ to be the
$U$-geometrical variable. However it seems more appealing to use a
combined expression $\tau$ involving $R$, $\rho$ and $\Theta$
given by
\begin{equation}
\tau = R + (1 + 3\lambda) \hspace{0.1cm}\rho - \frac{2}{3}
\hspace{0.1cm} \Theta^{2}. \protect\label{1239}
\end{equation}
In the unperturbed FRW background this quantity is defined in
terms of the curvature scalar of the 3-dimensional space and the
scale factor $A(t)$:
\begin{displaymath}
\frac{^{(3)}R}{A^{2}}.
\end{displaymath}
We define then the new associated variable $\tau_{\alpha}$ as
\begin{equation}
\tau_{\alpha} = {h_{\alpha}}^{\beta} \hspace{0.1cm}\tau_{,\beta}.
\protect\label{1311}
\end{equation}
This quantity $\tau_{\alpha}$ vanishes in the background. Its
perturbation can be written in terms of the previous variations,
since Einstein\rq s equations give
\begin{displaymath}
\tau = 2 \hspace{0.1cm}\left(\rho - \frac{1}{3} \hspace{0.1cm}
\Theta^{2}\right).
\end{displaymath}

\section{APPENDIX B: QUASI-MAXWELLIAN EQUATIONS}
\protect\label{apb}

We list below the quasi-Maxwellian equations of gravity. They are
obtained from Bianchi identities as true dynamical equations which
describe the propagation of gravitational disturbances. Making use
of Einstein\rq s equations and the definition of Weyl tensor,
Bianchi identities can be written in an equivalent form as
\begin{eqnarray}
{W^{\alpha\beta\mu\nu}}_{;\nu} & = & \frac{1}{2}R^{\mu[\alpha
;\beta]} -
\frac{1}{12}g^{\mu[\alpha}R^{,\beta]} \nonumber \\
& = & - \frac{1}{2}T^{\mu[\alpha ;\beta]} +
\frac{1}{6}g^{\mu[\alpha}T^{,\beta]}. \nonumber
\end{eqnarray}

Using the decomposition of Weyl tensor in terms of
$E_{\alpha\beta}$ and $H_{\alpha\beta}$ (see Section \ref{gauge})
and projecting appropriately, Einstein\rq s equations can be
written in a form which is similar to Maxwell\rq s equations.
There are 4 independent projections for the divergence of Weyl
tensor, namely:
\begin{displaymath}
\begin{array}{ll}
{W^{\alpha\beta\mu\nu}}_{;\nu} \hspace{0.1cm}V_{\beta} V_{\mu}
\hspace{0.1cm}
{h_{\alpha}}^{\sigma}, \\ \\
{W^{\alpha\beta\mu\nu}}_{;\nu} \hspace{0.1cm}
{\eta^{\sigma\lambda}}_{\alpha\beta} \hspace{0.1cm}V_{\mu}V_{\lambda}, \\ \\
{W^{\alpha\beta\mu\nu}}_{;\nu} \hspace{0.1cm}{h_{\mu}}^{(\sigma}
\hspace{0.1cm}
{\eta^{\tau)\lambda}}_{\alpha\beta} \hspace{0.1cm}V_{\lambda}, \\ \\
{W^{\alpha\beta\mu\nu}}_{;\nu} \hspace{0.1cm}V_{\beta}
\hspace{0.1cm} h_{\mu (\tau}h_{\sigma )\alpha}.
\end{array}
\end{displaymath}

The unperturbed quasi-Maxwellian equations are thus given by:
\begin{eqnarray}
h^{\varepsilon\alpha} h^{\lambda\gamma} \hspace{0.1cm}
E_{\alpha\lambda;\gamma} & + & {\eta^{\varepsilon}}_{\beta\mu\nu}
V^{\beta} \hspace{0.1cm}H^{\nu\lambda} \hspace{0.1cm}
{\sigma^{\mu}}_{\lambda} +
3H^{\varepsilon\nu} \hspace{0.1cm}\omega_{\nu} \nonumber \\
& = & \frac{1}{3} h^{\varepsilon\alpha}
\hspace{0.1cm}\rho_{,\alpha} + \frac{\Theta}{3} q^{\varepsilon} -
\frac{1}{2} ({\sigma^{\varepsilon}}_{\nu} -
3{\omega^{\varepsilon}}_{\nu})\hspace{0.1cm}
q^{\nu} \nonumber \\
& + & \frac{1}{2}\pi^{\varepsilon\mu} \hspace{0.1cm}a_{\mu} +
\frac{1}{2} h^{\varepsilon\alpha}
\hspace{0.1cm}{{\pi_{\alpha}}^{\nu}}_{;\nu} \protect\label{apb1}
\end{eqnarray}
\begin{eqnarray}
h^{\varepsilon\alpha} \hspace{0.1cm}h^{\lambda\gamma}
\hspace{0.1cm} H_{\alpha\lambda ;\gamma} & - &
{\eta^{\varepsilon}}_{\beta\mu\nu} \hspace{0.1cm}V^{\beta}
\hspace{0.1cm}E^{\nu\lambda} \hspace{0.1cm}
{\sigma^{\mu}}_{\lambda} - 3E^{\varepsilon\nu}
\hspace{0.1cm}\omega_{\nu}
\nonumber \\
& = & (\rho + p)\omega^{\varepsilon} - \frac{1}{2} \hspace{0.1cm}
\eta^{\varepsilon\alpha\beta\lambda}
\hspace{0.1cm}V_{\lambda} \hspace{0.1cm}q_{\alpha ;\beta} \nonumber \\
& + & \frac{1}{2}
\hspace{0.1cm}\eta^{\varepsilon\alpha\beta\lambda}
(\sigma_{\mu\beta} + \omega_{\mu\beta})
\hspace{0.1cm}{\pi^{\mu}}_{\alpha} \hspace{0.1cm}V_{\lambda}
\protect\label{apb2}
\end{eqnarray}
\begin{eqnarray}
{h_{\mu}}^{\varepsilon} {h_{\nu}}^{\lambda}
\hspace{0.1cm}\dot{H}^{\mu\nu} & + & \Theta H^{\varepsilon\lambda}
- \frac{1}{2}{H_{\nu}}^{(\varepsilon}
{h^{\lambda )}}_{\mu} \hspace{0.1cm}V^{\mu ;\nu} \nonumber \\
& + & \eta^{\lambda\nu\mu\gamma} \eta^{\varepsilon\beta\tau\alpha}
\hspace{0.1cm}V_{\mu} V_{\tau} \hspace{0.1cm}H_{\alpha\gamma}
\hspace{0.1cm}
\Theta_{\nu\beta} \nonumber \\
& - & a_{\alpha} {E_{\beta}}^{(\lambda}
\eta^{\varepsilon )\gamma\alpha\beta} \hspace{0.1cm}V_{\gamma} \nonumber \\
& + & \frac{1}{2}{{E_{\beta}}^{\mu}}_{;\alpha} \hspace{0.1cm}
{h_{\mu}}^{(\varepsilon}\eta^{\lambda )\gamma\alpha\beta}
\hspace{0.1cm}
V_{\gamma} \nonumber \\
& = & - \frac{3}{4}q^{(\varepsilon}\omega^{\lambda )} +
\frac{1}{2}h^{\varepsilon\lambda} \hspace{0.1cm}q^{\mu}
\omega_{\mu}
\nonumber \\
& + & \frac{1}{4}{\sigma_{\beta}}^{(\varepsilon}
\eta^{\lambda )\alpha\beta\mu} \hspace{0.1cm}V_{\mu} q_{\alpha} \nonumber \\
& + & \frac{1}{4} h^{\nu(\varepsilon}\eta^{\lambda
)\alpha\beta\mu} \hspace{0.1cm}V_{\mu}
\hspace{0.1cm}\pi_{\nu\alpha ;\beta} \protect\label{apb3}
\end{eqnarray}
\begin{eqnarray}
{h_{\mu}}^{\varepsilon} {h_{\nu}}^{\lambda}
\hspace{0.1cm}\dot{E}^{\mu\nu} & + & \Theta E^{\varepsilon\lambda}
- \frac{1}{2} {E_{\nu}}^{(\varepsilon}
{h^{\lambda )}}_{\mu} \hspace{0.1cm}V^{\mu ;\nu} \nonumber \\
& + & \eta^{\lambda\nu\mu\gamma} \eta^{\varepsilon\beta\tau\alpha}
\hspace{0.1cm}V_{\mu}V_{\tau} \hspace{0.1cm}E_{\alpha\gamma}
\Theta_{\nu\beta}\nonumber \\
&+& a_{\alpha} {H_{\beta}}^{(\lambda}
\eta^{\varepsilon )\gamma\alpha\beta} \hspace{0.1cm}V_{\gamma} \nonumber \\
& - & \frac{1}{2} {{H_{\beta}}^{\mu}}_{;\alpha} \hspace{0.1cm}
{h_{\mu}}^{(\varepsilon}\eta^{\lambda )\gamma\alpha\beta}
\hspace{0.1cm}
V_{\gamma} \nonumber \\
& = & \frac{1}{6}h^{\varepsilon\lambda} ({q^{\mu}}_{;\mu} -
q^{\mu} a_{\mu} - \pi^{\nu\mu} \sigma_{\mu\nu}) \nonumber \\
& - & \frac{1}{2}(\rho + p) \sigma^{\varepsilon\lambda} +
\frac{1}{2}
q^{(\varepsilon} a^{\lambda )} \nonumber \\
& - & \frac{1}{4} h^{\mu (\varepsilon} h^{\lambda )\alpha}
\hspace{0.1cm} q_{\mu ;\alpha} +
\frac{1}{2}{h_{\alpha}}^{\varepsilon}{h_{\mu}}^{\lambda}
\hspace{0.1cm}\dot{\pi}^{\alpha\mu} \nonumber \\
& + & \frac{1}{4}{\pi_{\beta}}^{(\varepsilon} \sigma^{\lambda
)\beta} - \frac{1}{4}{\pi_{\beta}}^{(\varepsilon} \omega^{\lambda
)\beta} + \frac{1}{6} \Theta\pi^{\varepsilon\lambda}.
\protect\label{apb4}
\end{eqnarray}

The contracted Bianchi identities and Einstein\rq s equations give
the conservation law
\begin{displaymath}
{T^{\mu\nu}}_{;\nu} = 0.
\end{displaymath}
Projecting it both in the parallel and the orthogonal subspaces we
obtain:
\begin{displaymath}
\begin{array}{ll}
{T^{\mu\nu}}_{;\nu} V_{\mu} = 0, \\ \\
{T^{\mu\nu}}_{;\nu} {h_{\mu}}^{\alpha} = 0,
\end{array}
\end{displaymath}
which give the following equations:
\begin{equation}
\dot{\rho} + (\rho + p)\Theta + \dot{q}^{\mu} V_{\mu} +
{q^{\alpha}}_{;\alpha} - \pi^{\mu\nu}\Theta_{\mu\nu} = 0,
\protect\label{apb5}
\end{equation}
\begin{eqnarray}
(\rho + p)a_{\alpha} & - & p_{,\mu}{h^{\mu}}_{\alpha} +
\dot{q}_{\mu} {h^{\mu}}_{\alpha} + \Theta q_{\alpha} \nonumber \\
& + & q^{\nu}\Theta_{\alpha\nu} + q^{\nu}\omega_{\alpha\nu} +
{{\pi_{\alpha}}^{\nu}}_{;\nu}  \nonumber \\
&+&\pi^{\mu\nu}\Theta_{\mu\nu}V_{\alpha} = 0, \protect\label{apb6}
\end{eqnarray}
and, from the definition of Riemann curvature tensor
\begin{displaymath}
V_{\mu ;\alpha ;\beta} - V_{\mu ;\beta ;\alpha} =
R_{\mu\varepsilon\alpha\beta} V^{\varepsilon},
\end{displaymath}
we obtain the equations of motion for the unperturbed kinematical
quantities as:
\begin{equation}
\dot{\Theta} + \frac{\Theta^{2}}{3} + 2\sigma^{2} + 2\omega^{2} -
{a^{\alpha}}_{;\alpha} = R_{\mu\nu} V^{\mu} V^{\nu},
\protect\label{apb7}
\end{equation}
\begin{eqnarray}
{h_{\alpha}}^{\mu} {h_{\beta}}^{\nu}
\hspace{0.1cm}\dot{\sigma}_{\mu\nu} & + &
\frac{1}{3}h_{\alpha\beta} (-2\omega^{2} - 2\sigma^{2} +\nonumber \\
{a^{\lambda}}_{;\lambda}) + a_{\alpha} a_{\beta} & - &
\frac{1}{2}{h_{\alpha}}^{\mu}{h_{\beta}}^{\nu} \hspace{0.1cm}
(a_{\mu ;\nu} + a_{\nu ;\mu})\nonumber \\
 & +& \frac{2}{3}\Theta\sigma_{\alpha\beta} +
\sigma_{\alpha\mu}{\sigma^{\mu}}_{\beta} +
\omega_{\alpha\mu}{\omega^{\mu}}_{\beta}
\nonumber \\
& = & R_{\alpha\varepsilon\beta\nu}V^{\varepsilon}V^{\nu} -
\frac{1}{3} R_{\mu\nu}V^{\mu}V^{\nu}h_{\alpha\beta},
\protect\label{apb8}
\end{eqnarray}
\begin{eqnarray}
{h_{\alpha}}^{\mu} {h_{\beta}}^{\nu}
\hspace{0.1cm}\dot{\omega}_{\mu\nu} & - &
\frac{1}{2}{h_{\alpha}}^{\mu}{h_{\beta}}^{\nu}(a_{\mu ;\nu} -
a_{\nu ;\mu}) + \frac{2}{3}\Theta\omega_{\alpha\beta} \nonumber \\
& + & \sigma_{\alpha\mu}{\omega^{\mu}}_{\beta} - \sigma_{\beta\mu}
{\omega^{\mu}}_{\alpha} = 0. \protect\label{apb9}
\end{eqnarray}

We also obtain from the definition of $R_{\alpha\beta\mu\nu}$
three constraint equations:
\begin{equation}
\frac{2}{3}\Theta_{,\mu}{h^{\mu}}_{\lambda} -
({\sigma^{\alpha}}_{\gamma} +
{\omega^{\alpha}}_{\gamma})_{;\alpha}{h^{\gamma}}_{\lambda} -
a^{\nu} (\sigma_{\lambda\nu} + \omega_{\lambda\nu}) =
R_{\mu\nu}V^{\mu} {h^{\nu}}_{\lambda}, \protect\label{apb10}
\end{equation}
\begin{equation}
{\omega^{\alpha}}_{;\alpha} + 2\omega^{\alpha}
\hspace{0.1cm}a_{\alpha} = 0, \protect\label{apb11}
\end{equation}
\begin{equation}
- \frac{1}{2} \hspace{0.1cm}{h_{\tau}}^{\varepsilon}
\hspace{0.1cm} {h_{\lambda}}^{\alpha}
\hspace{0.1cm}{\eta_{\varepsilon}}^{\beta\gamma\nu}
\hspace{0.1cm}V_{\nu} \hspace{0.1cm}(\sigma_{\alpha\beta} +
\omega_{\alpha\beta})_{;\gamma} + a_{(\tau}
\hspace{0.1cm}\omega_{\lambda )} = H_{\tau\lambda}.
\protect\label{apb12}
\end{equation}

These results constitute a set of 12 equations which will be used
to describe the evolution of small perturbations in FRW
background. Writing all the perturbed quantities in the form
\begin{displaymath}
X_{(perturbed)} = X_{(background)} + \delta X
\end{displaymath}
and after straightforward manipulations we finally obtain the
perturbed equations from the set of equations
(\ref{apb1})-(\ref{apb12}) as:
\begin{eqnarray}
(\delta E^{\mu\nu})^{\bullet} \hspace{0.1cm}
{h_{\mu}}^{\alpha}{h_{\nu}}^{\beta} & + & \Theta
\hspace{0.1cm}(\delta E^{\alpha\beta}) - \frac{1}{2} (\delta
{E_{\nu}}^{(\alpha}){h^{\beta )}}_{\mu}
\hspace{0.1cm}V^{\mu ;\nu} \nonumber \\
& + & \frac{\Theta}{3}\eta^{\beta\nu\mu\varepsilon} \hspace{0.1cm}
\eta^{\alpha\gamma\tau\lambda} \hspace{0.1cm}V_{\mu}V_{\tau}
(\delta E_{\varepsilon\lambda}) \hspace{0.1cm}h_{\gamma\nu} \nonumber \\
& - & \frac{1}{2} (\delta {{H_{\lambda}}^{\mu}})_{;\gamma}
\hspace{0.1cm}
{h_{\mu}}^{(\alpha}\eta^{\beta )\tau\gamma\lambda}V_{\tau} \nonumber \\
& = & - \frac{1}{2}(\rho + p)\hspace{0.1cm}(\delta
\sigma^{\alpha\beta})
\nonumber \\
& + & \frac{1}{6} \hspace{0.1cm}h^{\alpha\beta}
\hspace{0.1cm}(\delta q^{\mu})_{;\mu} - \frac{1}{4}
\hspace{0.1cm}h^{\mu (\alpha} h^{\beta )\nu}
\hspace{0.1cm}(\delta q_{\mu})_{;\nu} \nonumber \\
& + & \frac{1}{2} \hspace{0.1cm}h^{\mu(\alpha} h^{\beta )\nu}
\hspace{0.1cm} (\delta\Pi_{\mu\nu})^{\bullet} + \frac{1}{6}
\hspace{0.1cm}\Theta \hspace{0.1cm}(\delta\Pi^{\alpha\beta})\nonumber \\
\protect\label{apb13}
\end{eqnarray}
\begin{eqnarray}
(\delta H^{\mu\nu})^{\bullet} \hspace{0.1cm}
{h_{\mu}}^{\alpha}{h_{\nu}}^{\beta} & + & \Theta \hspace{0.1cm}
(\delta H^{\alpha\beta}) - \frac{1}{2} (\delta
{H_{\nu}}^{(\alpha})
{h^{\beta )}}_{\mu} \hspace{0.1cm}V^{\mu ;\nu} \nonumber \\
& + & \frac{\Theta}{3} \eta^{\beta\nu\mu\varepsilon}
\hspace{0.1cm} \eta^{\alpha\lambda\tau\gamma}
\hspace{0.1cm}V_{\mu}V_{\tau} \hspace{0.1cm}
(\delta H_{\varepsilon\gamma}) \hspace{0.1cm}h_{\lambda\nu} \nonumber \\
& - & \frac{1}{2} (\delta {{E_{\lambda}}^{\mu}})_{;\tau}
\hspace{0.1cm} {h_{\mu}}^{(\alpha}\eta^{\beta )\tau\gamma\lambda}
\hspace{0.1cm}V_{\gamma}
\nonumber \\
& = & \frac{1}{4} \hspace{0.1cm}h^{\nu (\alpha} \eta^{\beta
)\varepsilon\tau\mu} \hspace{0.1cm}V_{\mu} \hspace{0.1cm}
(\delta\Pi_{\nu\varepsilon})_{;\tau} \protect\label{apb14}
\end{eqnarray}
\begin{equation}
(\delta H_{\alpha\mu})_{;\nu} h^{\alpha\varepsilon} h^{\mu\nu} =
(\rho + p) \hspace{0.1cm}(\delta\omega^{\varepsilon}) -
\frac{1}{2} \hspace{0.1cm} \eta^{\varepsilon\alpha\beta\mu}
\hspace{0.1cm}V_{\mu} \hspace{0.1cm}(\delta q_{\alpha})_{;\beta}
\protect\label{apb15}
\end{equation}
\begin{eqnarray}
(\delta E_{\alpha\mu})_{;\nu} h^{\alpha\varepsilon}
\hspace{0.1cm}h^{\mu\nu} & = & \frac{1}{3} (\delta\rho)_{,\alpha}
h^{\alpha\varepsilon} - \frac{1}{3}
\dot{\rho} \hspace{0.1cm}(\delta V^{\varepsilon}) \nonumber \\
& - & \frac{1}{3} \hspace{0.1cm}\rho_{,0} \hspace{0.1cm}(\delta
V^{0})
\hspace{0.1cm}V^{\varepsilon} \nonumber \\
& + & \frac{1}{2} \hspace{0.1cm}{h^{\varepsilon}}_{\alpha}
\hspace{0.1cm} (\delta\Pi^{\alpha\mu})_{;\mu} + \frac{\Theta}{3}
\hspace{0.1cm}(\delta q^{\varepsilon}) \protect\label{apb16}
\end{eqnarray}
\begin{equation}
(\delta\Theta)^{\bullet} + \frac{2}{3}\Theta
\hspace{0.1cm}(\delta\Theta) - (\delta {a^{\alpha}})_{;\alpha} = -
\frac{(1 + 3\lambda)}{2} \hspace{0.1cm} (\delta\rho)
\protect\label{apb17}
\end{equation}
\begin{eqnarray}
(\delta\sigma_{\mu\nu})^{\bullet} & + & \frac{1}{3}h_{\mu\nu}
(\delta {a^{\alpha}})_{;\alpha} - \frac{1}{2}(\delta
a_{(\alpha})_{;\beta)} \hspace{0.1cm}{h_{\mu}}^{\alpha}
\hspace{0.1cm}{h_{\nu}}^{\beta}
\nonumber \\
& + & \frac{2}{3}\Theta \hspace{0.1cm}(\delta\sigma_{\mu\nu}) = -
(\delta E_{\mu\nu}) - \frac{1}{2} (\delta\Pi_{\mu\nu})
\protect\label{apb18}
\end{eqnarray}
\begin{equation}
(\delta\omega^{\mu})^{\bullet} + \frac{2}{3}\Theta \hspace{0.1cm}
(\delta\omega^{\mu}) = \frac{1}{2} \eta^{\alpha\mu\beta\gamma}
\hspace{0.1cm}(\delta a_{\beta})_{ ;\gamma}
\hspace{0.1cm}V_{\alpha} \protect\label{apb19}
\end{equation}
\begin{eqnarray}
\frac{2}{3}(\delta\Theta)_{,\lambda}
\hspace{0.1cm}{h^{\lambda}}_{\mu} & - & \frac{2}{3}\dot{\Theta}
\hspace{0.1cm} (\delta V_{\mu}) + \frac{2}{3}
\hspace{0.1cm}\dot{\Theta} \hspace{0.1cm}(\delta V^{0})
\hspace{0.1cm}
{\delta_{\mu}}^{0} \nonumber \\
& - & {(\delta {\sigma^{\alpha}}_{\beta} + \delta
{\omega^{\alpha}}_{\beta})}_{;\alpha} {h^{\beta}}_{\mu} = -
(\delta q_{\mu}) \protect\label{apb20}
\end{eqnarray}
\begin{equation}
(\delta {\omega^{\alpha}})_{;\alpha} = 0 \protect\label{apb21}
\end{equation}
\begin{equation}
(\delta H_{\mu\nu}) = - \frac{1}{2} \hspace{0.1cm}
{h^{\alpha}}_{(\mu} \hspace{0.1cm} {h^{\beta}}_{\nu )}
((\delta\sigma_{\alpha\gamma})_{;\lambda} +
(\delta\omega_{\alpha\gamma})_{;\lambda}) \hspace{0.1cm}
{\eta_{\beta}}^{\varepsilon\gamma\lambda}
\hspace{0.1cm}V_{\varepsilon} \protect\label{apb22}
\end{equation}
\begin{equation}
(\delta\rho)^{\bullet} + \Theta \hspace{0.1cm}(\delta\rho + \delta
p) + (\rho + p) \hspace{0.1cm}(\delta\Theta) + (\delta
q^{\alpha})_{;\alpha} = 0 \protect\label{apb23}
\end{equation}
\begin{eqnarray}
\dot{p} \hspace{0.1cm}(\delta V_{\mu}) & + & p_{,0} \hspace{0.1cm}
(\delta V^{0}) \hspace{0.1cm}{\delta_{\mu}}^{0} - (\delta
p)_{,\beta} \hspace{0.1cm}{h^{\beta}}_{\mu} + (\rho + p)
\hspace{0.1cm}(\delta a_{\mu})
\nonumber \\
& + & h_{\mu\alpha} (\delta q^{\alpha})^{\bullet} +  \frac{4}{3}
\Theta \hspace{0.1cm}(\delta q_{\mu}) + h_{\mu\alpha}
\hspace{0.1cm}(\delta \pi^{\alpha\beta})_{;\beta} = 0.\nonumber \\
\protect\label{apb24}
\end{eqnarray}

\vspace{2.0cm}

{\bf Acknowledgements} M. Novello and J. M. Salim would like to
thank CNPq for a grant.


\newpage

\begin{thebibliography}{99}
\bibitem{hawking} S. W. Hawking and G. F. Elis, The Large Scale
Structure of Space-Time (Cambridge University Press, Cambridge,
England, 1973).
\bibitem{Bardeen} J. Bardeen, in Proceedings of GRG, Tiflis, U.S.S>R., 1968.
\bibitem{Barrabes} C. Barrabes and V. P. Frolov, Phys. Rev. D 53,
3215 (1996).
\bibitem{Cabo} A. Cabo and E. Ayon-Beato, gr-qc/9704073, (1997).
\bibitem{Mars} M. Mars, M. M. Martin-Prats, and J. M. M. Senovila,
Classical Quantum Gravity 13, L51 (1996).
\bibitem{Tseytlin} A. A. Tseytlin, Phys. Lett. B 363, 223 (1995).
\bibitem{Cvetic} M. Cvetic, Phys. Rev. Lett. 71, 815 (1993).
\bibitem{Horne} J. H. Horne and G. T. Horowitz, Nucl. Phys. B368,
444 (1992).
\bibitem{Ayon2} E. Ayon-Beato and A. Garcia, Phys. Rev. Lett. 80,
5056 (1998).
\bibitem{Ayon3} E. Ayon-Beato and A. Garcia, Gen. Rel. Grav. 31,
629, (1999).
\bibitem{Ayon4} E. Ayon-Beato and A. Garcia, Phys. Lett. B464,25,
(1999).
\bibitem{deSitter}W. de Sitter, Proc. K. Ned. Akad. Wet. 19, 1217,
(1917).

\bibitem{NovelloS}M. Novello and J. M. Salim, Phys. Rev. D20, 377,
(1979); A. Saa, E. Gunzig, L. Brenig, V. Faraoni, T. M. Rocha
Filho, and A. Figueiredo, gr-qc/0012105, 2000.

\bibitem{Mukhanov}V. Mukhanov and R. Brandenberger, Phys. Rev,
Lett. 68, 1969, (1992).See also R. Brandenberger, V. Mukhanov and
A. Sornborger, Phys. Rev. D 48, 1629, (1993); R. Moessner and M.
Trodden,{\it ibid} 51, 2801, (1995).

\bibitem{NovelloS2}M. Novello, L. A. R. Oliveira, J. M. Salim and
E. Elbaz, Int. J. Mod. Phys. D1, 641 (1993).


\bibitem{Murphy}G. L. Murphy, Phys. Rev. D8, 4231 (1973); J. M.
Salim and H. P. de Oliveira, Acta Phys. Pol. B 19, 649, (1988).

\bibitem{Veneziano}G. Veneziano, hep-th/0002094, 2000; R.
Klippert, V. A. De Lorenci, M. Novello and J. M. Salim, Phys.
Lett. B 472, 27 (2000).

\bibitem{Nelson}J. Acacio de Barros, N. Pinto-Neto and M. A.
Sagioro-Leal, Phys. Lett. A 241, 229, (1998).

\bibitem{NovelloM} V. A. De Lorenci, R. Klippert, M. Novello and J. M. Salim,
Phys. Rev. D {\bf 65}, (2002), 063501.

\bibitem{Cartier}Cyril Cartier, Ruth Durrer, Edmund J. Copeland,
Phys. Rev. D 67, 103517,(2003).

\bibitem{novellomg8} M. Novello in Proceedings of the eight Marcel
Grossmann Meeting on General RElativity, Ed. Tsvi Piran, World
Scientific, 1999.
\bibitem{Kolb} E. W. Kolb and M. S. Turner, in
{\em The Early Universe} (Addison Wesley, California, 1990).
\bibitem{brande}
R. Brandenberger, in the proceedings {\em VIII Brazilian School of
Cosmology and Gravitation}, ed.\ M. Novello (Editions Frontieres,
Singapore, 1996).
\bibitem{nelson} P. Peter and N. Pinto-Neto, Phys. Rev D vol 65,
023513 (2001)
\bibitem{footnote1}From a mathematical point of view, a
negative energy could also allow for a bouncing. We will not
examine this possibility in the present paper.
\bibitem{footnote2}If
we consider that the origin of these corrections come from quantum
fluctuations then the value of the constants $\alpha$ and $\beta$
are fixed by the calculations made by Heisenberg and Euler.
\bibitem{footnote3}Note that once we are dealing
with linear process each mode can be analysed separately.

\bibitem{Dunne}
G. Dunne and T. Hall, {\em Phys.\ Rev.\ D} {\bf 58}, 105022
(1998); G. Dunne, {\em Int.\ J. Mod.\ Phys.\ A} {\bf 12} (6), 1143
(1997).
\bibitem{Tajima}
T. Tajima, S. Cable, K. Shibata, and R. M. Kulsrud, {\em
Astrophys.\ J.} {\bf 390}, 309 (1992);
M. Giovannini and M. Shaposhnikov, {\em Phys.\ Rev.\ D} {\bf 57}
(4), 2186 (1998).
\bibitem{Campos}
A. Campos and B. L. Hu, {\em Phys.\ Rev.\ D} {\bf 58}, 125021
(1998).
\bibitem{Novello1}M. Novello, J.M. Salim, M.C. Motta da Silva, S.E.
Jorás and R. Klippert, Phys. Rev. D {\bf 51}, 450 (1995).
\bibitem{Novello2}M. Novello, J.M. Salim,M.C. Motta da Silva, S.E.
Jorás and R. Klippert, Phys. Rev. D {\bf 52}, 730 (1995).
\bibitem{Novello3}M. Novello, J.M. Salim,M.C. Motta da Silva and R. Klippert,
 Phys. Rev. D {\bf 54}, 2578 (19956).
 \bibitem{Hawking}S. W. Hawking, Ap. J. {\bf 145} (1966), 544.
 \bibitem{Lifshitz}E. M. Lifshitz and I.M. Khalatnikov, Adv.Phys. {\bf 12},
(1963), 185.
\bibitem{Ellis} G.F.R.Ellis and M.Bruni, Phys Rev D 40,6 (1989)
1804.
\bibitem{Landau}L. D. Landauand E. M. Lifshitz, {\bf Fluid Mechanics}.
Pergamon Press, N.Y., 1982.
\bibitem{Abramowitz} M. Abramowitz and Irene A. Stegun, {\bf Hand Book
of Mathematical Functions}. Dover Publications, N.
Y.,pg504,(1974).
\bibitem{footnote4}We
remind the reader that we restrain here our examination to
irrotational perturbation. The formulas which we obtain are thus
simpler. However the method of our analysis is not restrictive and
the study of generic cases can be obtained through the same lines.
\end{thebibliography}
\end{document}